\documentclass{aastex}          
\usepackage{spr-astr-addons}    
\usepackage{cases}
\usepackage{subfigure}
\usepackage{sidecap}

\newcommand{\etal}{{\it et al.}}
\begin{document}

\title{North--South Asymmetry in Solar Activity and  Solar Cycle Prediction,
V: Prediction for the North--South Asymmetry in the Amplitude of Solar Cycle~25}

\shorttitle{Prediction for the North--South Asymmetry in Solar Cycle~25}
\shortauthors{J. Javaraiah}

\author{J. Javaraiah\altaffilmark{1}}

\email{jajj55@yahoo.co.in; jdotjavaraiah@gmail.com} 

\altaffiltext{1}{
Bikasipura, Bengaluru-560 111, India \\
{Formerly  with Indian Institute of Astrophysics, Bengaluru-560 034, India}}

\begin{abstract}
There exists a small but statistically significant north--south asymmetry in
 most of the solar activity indices and it has  important implications on the
 solar dynamo mechanism.  
 Here we analyzed the daily sunspot-group  data  reported by the  Greenwich
 Photoheliographic Results (GPR) during the period 1874\,--\,1976,
 Debrecen Photoheligraphic Data (DPD) during the period 1977\,--\,2017, 
and the revised Version-2 of international sunspot number (ISSN) during 
the period 1874\,--\,2017. 
We determined the amplitudes (the largest 13-month smoothed monthly ISSN)
 of  Solar Cycles 12\,--\,24 and
 the 13-month smoothed monthly mean corrected areas of the sunspot
 groups in the Sun's whole-sphere (WSGA), northern hemisphere (NSGA), and
southern hemisphere (SSGA) at the epochs of the maxima of  Solar Cycles
 12\,--\,24. Using all these we obtained the relations  similar to  
that found in our earlier analyzes--$i.e.$ the existence of a high correlation  
between the sum of the areas of sunspot groups in the southern-hemisphere 
near-equatorial band during a small (7--9 months) interval just
 after a maximum epoch of
a solar cycle  and the  amplitude of next solar cycle--separately
 for  the Sun's whole-sphere and northern- and southern-hemispheres.
By using these relations  we predict  
$\approx$701 msh (millionth of solar hemisphere),   $\approx$429 msh, 
and $\approx$366 msh for
 the values of  WSGA,  NSGA, and SSGA, respectively, at the maximum
 epoch of  Solar Cycle~25.  We  predict $86 \pm 18$ for  the amplitude
 of Solar Cycle~25.
The 13-month smoothed monthly mean sunspot-group area highly 
correlate with that  of ISSN. 
 Using this relation and the predicted values of WSGA,  NSGA, and SSGA
 we obtain $68 \pm 11$  for the amplitude of Solar Cycle~25, which is 
slightly lower than the aforementioned predicted value,  and
$39 \pm 4$ and $31 \pm 6$ for the values of 
northern- and southern-hemispheres' sunspot numbers at the maximum  
epoch of Solar Cycle~25. The difference between the predicted 
 NSGA and SSGA and also that between northern- and  
southern-hemispheres'  sunspot numbers  at the maximum epoch of Cycle~25  
are considerably small. Overall, our results 
suggest that the amplitude of Cycle~25 would be 25\,\%--40\,\% smaller, 
and the corresponding north--south asymmetry  would be
much  smaller, than those of Cycle~24.  
\end{abstract}


\section{Introduction}
Prediction of the  strength of a solar cycle well in advance is important
in view of the effects of solar activity on the  terrestrial atmosphere  and
 the space environment.
There exists a small but statistically
 significant north--south asymmetry in most of the solar activity 
indices, such as  sunspots, solar flares, prominence eruptions, 
 coronal mass ejections, {\it etc.} \citep[see][]{nort14,hath15}.   
During the late Maunder minimum the north--south asymmetry was  
very large~\citep{soko94}. Asymmetric polar field reversals
 may be  a consequence of the asymmetry of solar activity~\citep{sk13}. 
North--south asymmetry in solar activity  is also known to vary on 
many time scales 
 \citep{vizo90,carb93,verma93,dd96,jg97,knaack04,chow13,rj15,chow16,mb16,deng16,jj19}. 
 There is also a suggestion on some systematic  phase shift in the activity 
cycle between the hemispheres
\citep[$e.g.$][]{zolo10,nort10,mura12,mci13}.
 However, in an our earlier analysis 
it is found that the cycle-maximum  is not always occurring first in one
 particular hemisphere~\citep{jj19}. 
 The physical process responsible for 
the north--south asymmetry in solar activity as well as its relation to 
variations in solar activity  is not clear yet 
\citep{vm17,badal17,sc18,nep19}.  
 Recently, it is suggested that there could be influence of some
 specific configurations 
of the giant planets in the origin of the well-known $\approx$12-year 
and 40\,--\,50-year periodicities in the north--south asymmetry of 
solar activity \citep{jj20}.
 North--south asymmetry in solar activity seems to
have some important implications on the solar dynamo process
\citep[][and references therein]{soko94,bd13,shetye15,nort14}. 
It seems the variations of the
 interplanetary magnetic field and the geomagnetic indices 
 more closely relate with the variation  of the north--south  asymmetry of the 
solar activity than with the variation of the global activity \citep{murak19}. 
Therefore, besides  prediction of the amplitude of a solar cycle,  prediction
 of the corresponding north--south asymmetry is also important
 for better understanding the basic processes of solar cycle and
the solar-Terrestrial relationship. Attempts for predicting 
the north--south asymmetry based on solar dynamo models have been also made
 \citep{dgd07,gc09}. 
\cite{hath16} and \cite{uh18} using an  Advective Flux Transport  code found
  an evidence
that during Solar Cycle~25 southern hemisphere will be  more active than
 northern hemisphere.

In our earlier
 analyzes 
\citep[][$i.e.$ Article~I, Article~II, Article~III]{jj07,jj08,jj15} it was
 found that the 
sum of the areas of the 
sunspot groups in  $0^\circ - 10^\circ$  latitude interval of  
the Sun's southern hemisphere during the interval 1.0 to 1.75 years after the 
maximum epoch of a solar cycle  ($i.e.$ around the time of 
polarities change in the Sun's global magnetic fields) correlate well with 
 the amplitude of the next cycle. By using this relation  the amplitudes of
 Solar Cycles 24 
\citep{jj07,jj08} and 25 \citep{jj15, jj17} were predicted. Recently, we 
analyzed the combined
 GPR and DPD daily sunspot group data during the period 1874\,--\,2017 
and determined the 13-month smoothed monthly   
 mean corrected areas of the sunspot groups in the Sun's  whole-sphere  (WSGA), 
  northern hemisphere (NSGA), and southern hemisphere  (SSGA).
 From these 
smoothed time series we had obtained  the values of the maxima and minima of 
the WSGA-, NSGA-, and SSGA-Cycles 12\,--\,24.  
We have predicted the lengths of
the WSGA-Cycles (and also the lengths of corresponding sunspot cycles) 24, 25,
 and 26 \citep[][$i.e.$ Article~IV]{jj19}.

In the present article  our main
 aim is to make a prediction for 
the north--south  asymmetry correspond to the amplitude of the
 upcoming Solar Cycle~25. In all our aforementioned earlier analyzes 
we had used the Version-1 of ISSN. Here we have used the revised 
Version-2 of ISSN.

In the next section we describe the data and analysis, in Section~3 we present 
the results, and in Section~4 we present the conclusions and discuss them 
briefly.

\section{Data and Analysis}
We have used here the same dataset of  sunspot-groups  that was used  
in \cite{jj19}. 
That is, we have analyzed the   combined Greenwich and DPD sunspot-group  
daily data  during the period April 1874\,--\,June 2017 that were 
downloaded from {\sf fenyi.solarobs.unideb.hu/pub/DPD/}. These data contain 
the values of the date and the time of observation, corrected whole-spot
area  ($A$ in msh:  millionth of solar hemisphere), 
heliographic latitude ($\lambda$) and longitude ($L)$, central 
meridian distance ($CMD$), $etc.$  of  sunspot groups~\citep[for details 
see][]{gyr10,bara16,gyr17}.  In \cite{jj19} 
besides determining the  actual amplitudes of WSGA-, 
NSGA-, and SSGA-Cycles 12\,--\,24, we 
have also determined   the values of WSGA, NSGA, 
and SSGA at the epochs of the maxima (maximum values of the 13-month 
smoothed monthly mean ISSN, $i.e.$ the values of $R_{\rm M}$)
  of the Sunspot Cycles 12\,--\,24.
 Hereafter  the values of 
WSGA, NSGA, and SSGA  at the epoch of $R_{\rm M}$ of a sunspot cycle  
 are  denoted  as $A_{\rm W}$, $A_{\rm N}$, and $A_{\rm S}$, 
respectively.
In that earlier article the values of $R_{\rm M}$
and the corresponding epoch
of each of the Sunspot Cycles 12\,--\,24 were determined by using 
the updated time series  of 
 the  13-month smoothed monthly mean Version-1 ISSN  during
the period 1874\,--\,2015. It  was downloaded from 
{\sf www.sidc.be/silso/datafiles}. Here we have used the 
 corresponding Version-2 SN time series during the period  
October 1874\,--\,June 2017 (used the file SN\_ms\_tot\_v2.0.txt). 
We have also used the Version-2 SN time series of  northern- and
 southern-hemispheres during the period
 July 1992\,--\,June 2017 (used the file SN\_ms\_hem\_V2.0.txt). 
All these are also downloaded from {\sf www.sidc.be/silso/datafiles}.
The details of changes and corrections in Version-2 SN are given in 
\citep{clette16}. Using the Version-2 SN time series we 
  determined the values of $R_{\rm M}$
 and the corresponding epochs of Sunspot Cycles 12\,--\,24~\citep[which are
 exactly the same as the values reported in][]{pesnell18}.
 Using these values we have determined the values of $A_{\rm W}$, $A_{\rm N}$,
 and $A_{\rm S}$. 
Most of the  values of $R_{\rm M}$ that are determined from the 
Version~2 of ISSN are different 
 from those determined from the Version-1 of ISSN and 
a few of the corresponding epochs are also slightly  different. 
Hence, at these epochs  
 the values of $A_{\rm W}$, $A_{\rm N}$, and $A_{\rm S}$ are 
slightly different from those of the corresponding parameters   
  of the whole sphere, northern hemisphere, and 
southern hemisphere, respectively, that are given in Table~1 of \cite{jj19}. 
In the cases of only a few cycles the values of $A_{\rm W}$, $A_{\rm N}$, and 
$A_{\rm S}$ represent the respective maximum values of the area cycles 
\citep{jj19}.    

Here we have used  
 the same method that was used in our earlier articles 
for predicting the amplitudes of  Solar Cycles 24 and 25
 \citep{jj07, jj08, jj15}. 
 We determined 
 the correlations of the values of  $A_{\rm W}$, $i.e.$ the values of 
WSGA at the maximum epochs of Cycles 13\,--\,24 
with  the  sums of the areas of the
sunspot groups in   $0^\circ - 10^\circ$ latitude interval of
the Sun's southern hemisphere during the different time intervals that are 
close to the maximum epochs of  Cycles 12\,--\,23. We choose the
 time intervals in which the sum of the area  has a maximum correlation 
with $A_{\rm W}$ and then we obtained the linear best-fit of 
 the area-sum and $A_{\rm W}$. 
 By using the obtained linear relationship and  the area-sum in the
 time interval that is close 
to the maximum epoch of Cycle 24, we predicted the value of
 $A_{\rm W}$ of Cycle 25.  
 Similarly, we predicted the values of  
 $A_{\rm N}$ and $A_{\rm S}$, 
  $i.e.$ the values of  NSGA and SSGA  
 at the maximum epoch of  Solar Cycle 25.
It should be noted  that 
the values of the amplitudes and their epochs of some of the cycles are 
different from the corresponding values that were taken from 
{\sf www.ngdc.noaa.gov} and  were  
used in our earlier analyzes \citep{jj07,jj08,jj15,jj17}. 
 We have already predicted the 
amplitude ($R_{\rm M}$) of Cycle~25 \citep{jj15,jj17}. However, 
since we have used here a different set of  the maximum values and the
corresponding epochs of Solar Cycles 12\,--\,24, hence, here   
we have also predicted the $R_{\rm M}$ of Sunspot Cycle~25. 
For the sake of reducing the uncertainty  
in the determined sums of the areas of sunspot groups due
to the foreshortening effect 
(if any) in the measured areas of sunspot groups,  we have not 
used the sunspot-group data correspond to the $|CMD| > 75^\circ$.
By using the predicted values  of
 $A_{\rm N}$ and $A_{\rm S}$ of Cycle~25, 
we determined the corresponding  north--south asymmetry, 
$\frac{A_{\rm N} - A_{\rm S}}{A_{\rm N} + A_{\rm  S}}$. In the next section 
we described the results.

Here it may be worth to mention that the (linear) fits that correspond
 to the poorer correlations are less accurate. Hence, obviously, the predictions
 are less accurate and unreliable. In fact, even in the case of the 
 predictions based on the
 polar fields at minimum of cycle could more uncertain and fail if the polar
 fields were used early before the start of the cycle. Therefore, we have
 found the exact interval through partitioning the datasets until a maximum
 correlation is found. The sunspot-group data near the maximum of a solar
 cycle is well measured and the sizes of the intervals that yielded the 
 maximum correlations are still sufficiently large (7 months and above).
In addition, these intervals may have some physical significance because they 
 comprised the epochs of polarities change in the Sun's global magnetic fields.

For the sake of the readers convenience we listed below the
 meanings of all the abbreviations that are used here, because  they  
 are large. 
\begin{itemize}
\item SN - 13-month  smoothed monthly mean Version-2 ISSN 
during  the period October 1874\,--\,June 2017,
\item ${\rm SN}_{\rm N}$ - 13-month  smoothed monthly mean  Version-2 ISSN 
 in northern hemisphere during the period July 1992\,--\,June 2017,
\item ${\rm SN}_{\rm S}$ - 13-month  smoothed monthly mean Version-2 ISSN
  in southern  hemisphere during  the period July 1992\,--\,June 2017,
\item $n$ - Waldmeier solar cycle number,
\item $T_{\rm M}$ - maximum epoch of a solar cycle,
\item $R_{\rm M}$ - value of SN  at $T_{\rm M}$,
\item $\sigma_{\rm R}$ - error in $R_{\rm M}$, 
\item $A_{\rm W}$ - the value of the 13-month smoothed monthly mean 
 area of the sunspot groups in whole sphere (WSGA)  at $T_{\rm M}$,
\item $A_{\rm N}$ - the value of the 13-month smoothed monthly mean
 area of the sunspot groups in northern hemisphere (NSGA) at $T_{\rm M}$,
\item $A_{\rm S}$ - the value of the 13-month smoothed monthly mean 
 area of the sunspot groups in southern hemisphere (SSGA) at $T_{\rm M}$,
\item $T^*_{\rm M}$ - the time-interval of $1.15$ year to $1.75$ year just
 after $T_{\rm M}$ of a cycle,
\item $T^*_{\rm W}$ - the time-interval of $0.95$ year to $1.6$ year just
 after $T_{\rm M}$ of a cycle,
\item $T^*_{\rm N}$ - the time-interval of $0.95$ year to $1.75$ year just
 after $T_{\rm M}$ of a cycle,
\item $T^*_{\rm S}$ - the time-interval of $0.75$ year to $1.65$ year just
 after $T_{\rm M}$ of a cycle,
\item $A^*_{{\rm R}}$ - the sum of the areas of sunspot groups in
 $0^\circ - 10^\circ$ latitude interval of the southern hemisphere
 during $T_{\rm M}^*$ of  Cycle $n$, which has maximum correlation 
 with $R_{\rm M}$ of Cycle $n + 1$,
\item $A^*_{{\rm W}}$ - the sum of the areas of sunspot groups in
 $0^\circ - 10^\circ$ latitude interval of the southern hemisphere
 during $T^*_{\rm W}$ of  Cycle $n$, which has maximum correlation 
with $A_{\rm W}$ of Cycle $n + 1$,
\item $A^*_{{\rm N}}$ - the sum of the areas of sunspot groups in
 $0^\circ - 10^\circ$ latitude interval of the southern hemisphere
 during $T^*_{\rm N}$ of  Cycle $n$, which has maximum correlation with
 $A_{\rm N}$ of Cycle $n + 1$,
\item $A^*_{{\rm S}}$ - the sum of the areas of sunspot groups in
 $0^\circ - 10^\circ$ latitude interval of the southern hemisphere
 during $T^*_{\rm S}$ of Cycle $n$, which has maximum correlation with
 $A_{\rm S}$ of Cycle $n + 1$,
\item $T_{\rm m}$ - the  preceding  minimum epoch of a  cycle,
\item $T^*_{\rm m}$ - the time interval around $T_{\rm m}$ of a cycle, and
\item $a^*_{\rm R}$ - the sum of the areas of sunspot groups in
 $0^\circ - 10^\circ$ latitude interval of the northern hemisphere
during $T^*_{\rm m}$ of  Cycle $n$, which has maximum correlation 
with $R_{\rm M}$ of Cycle $n + 1$.
\end{itemize}

\section{Results}
Fig.~\ref{fig1} shows the cycle-to-cycle variations in $A_{\rm W}$, $A_{\rm N}$,
 $A_{\rm S}$, and $R_{\rm M}$.
As can be seen in this figure, there exists a considerable difference 
between the variations in $A_{\rm N}$ and $A_{\rm S}$. The correlation 
between $A_{\rm N}$ and $A_{\rm S}$ is found  considerably weak 
(correlation coefficient $r = 0.45$). However, the correlation ($r = 0.85$) 
between $A_{\rm N}$ and $A_{\rm W}$, 
and that  ($r = 0.86$) between $A_{\rm S}$ and $A_{\rm W}$, are found to 
be almost the same.   In Table~\ref{table1} we have given the values of 
 $A^*_{\rm R}$, $A^*_{\rm W}$,   $A^*_{\rm N}$, and  $A^*_{\rm S}$, $viz.$, 
 the sums of the areas of the sunspot groups in $0^\circ - 10^\circ$
latitude interval of the southern hemisphere  during  the 
intervals $T^*_{\rm M}$,  $T^*_{\rm W}$,  $T^*_{\rm N}$, and  $T^*_{\rm S}$, 
 respectively, 
just after $T_{\rm M}$ of a solar cycle when each of the aforementioned
  sums of the areas of
 the sunspot groups in a Cycle $n$ well correlate    
 with the corresponding   
 $R_{\rm M}$, $A_{\rm W}$,   $A_{\rm N}$, and  $A_{\rm S}$ of  Cycle $n+1$.
The intervals  $T^*_{\rm M}$,  $T^*_{\rm W}$,  $T^*_{\rm N}$, 
and $T^*_{\rm S}$ started from the years 1.15, 0.95, 0.95, 
and 0.75, respectively, and ended at the years 1.75, 1.6, 1.75, and 1.65, 
respectively, from $T_{\rm M}$ of a cycle
($T^*_{\rm S}$ started and also ended slightly earlier than
 $T^*_{\rm N}$). It should be noted  that, as already mentioned above,
 the values of $R_{\rm M}$, and those of $T_{\rm M}$ of some of the 
solar cycles,  
that are used here  differ with those were used in the earlier
 analyzes~\citep{jj07,jj08,jj15,jj17}.  
Therefore, 
the intervals  $T^*_{\rm M}$,  $T^*_{\rm W}$,  $T^*_{\rm N}$, 
and $T^*_{\rm S}$
 of the present analysis differ with the corresponding earlier ones. However, 
each of these intervals  of the present analysis 
 comprised the corresponding  earlier ones in large extent.
 The results have been specifically optimized to produce the maximum 
correlation. In Figure~11 of Article~III, we have shown that how sensitive 
the results are 
at small variations in the exact start and stop  times of the intervals 
over which the sums of the areas of sunspot-groups are made. 
There is a considerable consistency in the data sampling.  That is, 
the intervals  $T^*_{\rm M}$,  $T^*_{\rm W}$,  $T^*_{\rm N}$, 
and $T^*_{\rm S}$ are sufficiently long. Hence, small changes  
in the  start and stop times of these intervals have no significant 
change in the results.

\begin{figure}[ht]
\centering
\includegraphics[width=8.0cm]{nsp-fig1.eps}
\caption{Cycle-to-cycle variations in
$A_{\rm W}$   ({\large $\bullet{- -}\bullet$}   {\it curve}),
 $A_{\rm N}$ ($\blacksquare$---$\blacksquare$ {\it curve}), and 
$A_{\rm S}$ ({\bf $\square$$\cdots$$\square$} {\it curve}), $viz.$ the 13-month 
smoothed monthly mean areas of the sunspot groups in the Sun's whole sphere, 
northern-, and southern-hemispheres, respectively,  at the epoch of 
$R_{\rm M}$, $i.e.$ at the largest  13-month smoothed monthly mean ISSN 
 of a solar cycle. The {\bf \tiny $\bigcirc${\scriptsize- -}$\bigcirc$} 
 {\it curve} represents 
the variation in $R_{\rm M}$. The corresponding mean values of all these 
quantities are shown with horizontal lines.
 The errors in $R_{\rm M}$ are 
also shown. The errors in the areas of sunspot groups are not available.}
\label{fig1}
\end{figure}

\begin{table}
\centering
\caption{$R_{\rm M}$ represents the maximum (the largest 13-month smoothed 
monthly mean ISSN) and $T_{\rm M}$  is the 
corresponding  epoch (year) of a Solar Cycle $n$.
  $A_{\rm W}$,   $A_{\rm N}$, and  $A_{\rm S}$ represent the  
  13-month smoothed monthly mean areas 
(msh) of the sunspot groups in the Sun's  whole-sphere,   
northern hemisphere, and southern hemisphere, respectively,  
 at  $T_{\rm M}$  of a solar cycle. 
 $A^*_{\rm R}$,   $A^*_{\rm W}$,   $A^*_{\rm N}$, and  $A^*_{\rm S}$ 
represent  the sums of the areas (msh) of the sunspot groups  
(normalized by 1000) in  $0^\circ-10^\circ$ latitude intervals of the 
southern hemisphere  during  the time intervals $T^*_{\rm M}$, 
 $T^*_{\rm W}$,  $T^*_{\rm N}$, and  $T^*_{\rm S}$,  respectively, 
just after $T_{\rm M}$ of a solar cycle.
 $\sigma_{\rm R}$ represents  the error in $R_{\rm M}$. Errors in the 
areas of sunspot groups are not available.} 

{\tiny 
\begin{tabular}{lccccc}
\hline
  \noalign{\smallskip}
$n$&$T_{\rm M}$&$R_{\rm M}$& $\sigma_{\rm R}$&$T^*_{\rm M}$&$A^*_{\rm R}$\\
  \noalign{\smallskip}
\hline
  \noalign{\smallskip}
12&1883.958&  124.4&12.5&1885.11-1885.71&  30.91\\
13&1894.042&  146.5&10.8&1895.19-1895.7&  27.02\\
14&1906.123&  107.1& 9.2&1907.27-1907.87&  32.34\\
15&1917.623&  175.7&11.8&1918.77-1919.37&  32.48\\
16&1928.290&  130.2&10.2&1929.44-1930.04&  70.20\\
17&1937.288&  198.6&12.6&1938.44-1939.04&  71.62\\
18&1947.371&  218.7&10.3&1948.52-1949.12& 103.85\\
19&1958.204&  285.0&11.3&1959.35-1959.95&  31.67\\
20&1968.874&  156.6& 8.4&1970.02-1970.62&  72.58\\
21&1979.958&  232.9&10.2&1981.11-1981.71&  81.31\\
22&1989.874&  212.5&12.7&1991.02-1991.62&  55.36\\
23&2001.874&  180.3&10.8&2003.02-2003.62&  30.50\\
24&2014.288&  116.4& 8.2&2015.44-2016.04&   6.20\\
\\

&&$A_{\rm W}$&&$T^*_{\rm W}$&$A^*_{\rm W}$\\
  \noalign{\smallskip}
12&1883.958&   1370&&1884.91-1885.56&  38.74\\
13&1894.042&   1616&&1894.99-1895.64&  32.14\\
14&1906.123&   1043&&1907.07-1907.72&  32.32\\
15&1917.623&   1535&&1918.57-1919.22&  33.21\\
16&1928.290&   1324&&1929.24-1929.89&  60.21\\
17&1937.288&   2119&&1938.24-1938.89&  77.47\\
18&1947.371&   2641&&1948.32-1948.97&  96.41\\
19&1958.204&   3441&&1959.15-1959.80&  36.25\\
20&1968.874&   1556&&1969.82-1970.47&  56.32\\
21&1979.958&   2121&&1980.91-1981.56&  56.19\\
22&1989.874&   2268&&1990.82-1991.47&  64.09\\
23&2001.874&   2157&&2002.82-2003.47&  32.25\\
24&2014.288&   1599&&2015.24-2015.89&   9.87\\
\\

&&$A_{\rm N}$&&$T^*_{\rm N}$&$A^*_{\rm N}$\\
  \noalign{\smallskip}
12&1883.958&    413&&1884.91-1885.71&  43.82\\
13&1894.042&    621&&1894.99-1895.79&  34.98\\
14&1906.123&    761&&1907.07-1907.87&  37.80\\
15&1917.623&    828&&1918.57-1919.37&  37.14\\
16&1928.290&    630&&1929.24-1930.04&  86.30\\
17&1937.288&   1308&&1938.24-1939.04&  94.14\\
18&1947.371&   1051&&1948.32-1949.12& 133.46\\
19&1958.204&   1748&&1959.15-1959.95&  43.10\\
20&1968.874&    951&&1969.82-1970.62&  76.58\\
21&1979.958&   1063&&1980.91-1981.71&  97.83\\
22&1989.874&   1120&&1990.82-1991.62&  69.79\\
23&2001.874&   1073&&2002.82-2003.62&  43.02\\
24&2014.288&    419&&2015.24-2016.04&  12.29\\
\\

&&$A_{\rm S}$&&$T^*_{\rm S}$&$A^*_{\rm S}$\\
  \noalign{\smallskip}
12&1883.958&    956&&1884.71-1885.61&  54.72\\
13&1894.042&    994&&1894.79-1895.69&  37.45\\
14&1906.123&    282&&1906.87-1907.77&  33.93\\
15&1917.623&    706&&1918.37-1919.27&  41.20\\
16&1928.290&    693&&1929.04-1929.94&  87.68\\
17&1937.288&    810&&1938.04-1938.94&  93.25\\
18&1947.371&   1590&&1948.12-1949.02& 114.50\\
19&1958.204&   1692&&1958.95-1959.85&  52.97\\
20&1968.874&    605&&1969.62-1970.52&  76.37\\
21&1979.958&   1057&&1980.71-1981.61&  86.56\\
22&1989.874&   1147&&1990.62-1991.52&  89.66\\
23&2001.874&   1084&&2002.62-2003.52&  89.80\\
24&2014.288&   1180&&2015.04-2015.94&  23.73\\
\hline
  \noalign{\smallskip}

\end{tabular}
}
\label{table1}
\end{table}

\begin{figure}[ht]
\centering
\includegraphics[width=8.0cm]{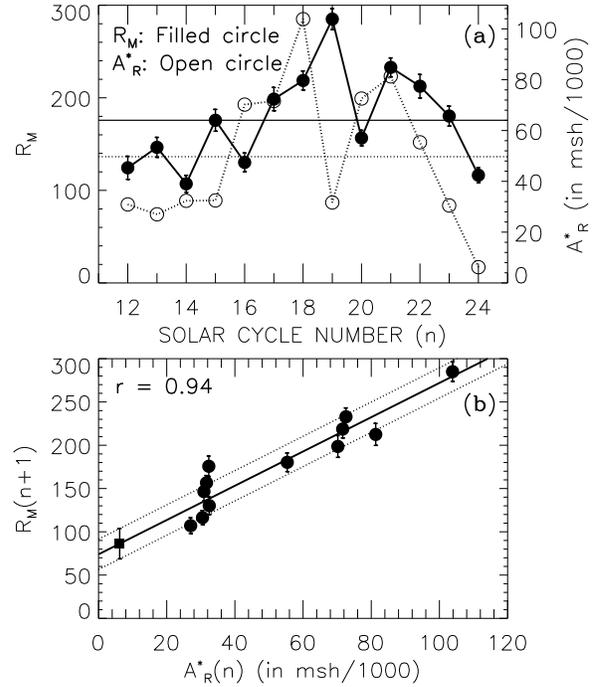}
\caption{(a) Plot of the amplitude  ($R_{\rm M}$), 
$i.e.$ the largest 13-month smoothed monthly mean ISSN  of
 a solar cycle ({\large $\bullet{\rm-}\bullet$} {\it curve})--and
$A^*_{\rm R}$,  $i.e.$ the sum of the areas of the sunspot groups  
in $0^\circ - 10^\circ$ latitude interval of the Sun's southern hemisphere 
during the interval $T^*_{\rm M}$ just after  the maximum epoch $T_{\rm M}$  
of a solar cycle ({\tiny $\bigcirc\cdots\bigcirc$} 
{\it curve})--versus the solar cycle number.
(b) The scatter plot of $A^*_{\rm R}$ of a Solar Cycle $n$  versus 
 $R_{\rm M}$  of Solar Cycle $n+1$. The {\it continuous line}  
represents  the corresponding linear best-fit and  
 the {\it dotted lines}  are drawn at one-rmsd
 ({\it root-mean-square  deviation}) level.  The values of the correlation
 coefficient ($r$) is also shown. 
 The $\blacksquare$ represents 
the predicted value $86 \pm 18$ for $R_{\rm M}$ of Cycle~25.}
\label{fig2}
\end{figure}

\begin{figure}[ht]
\centering
\includegraphics[width=8.0cm]{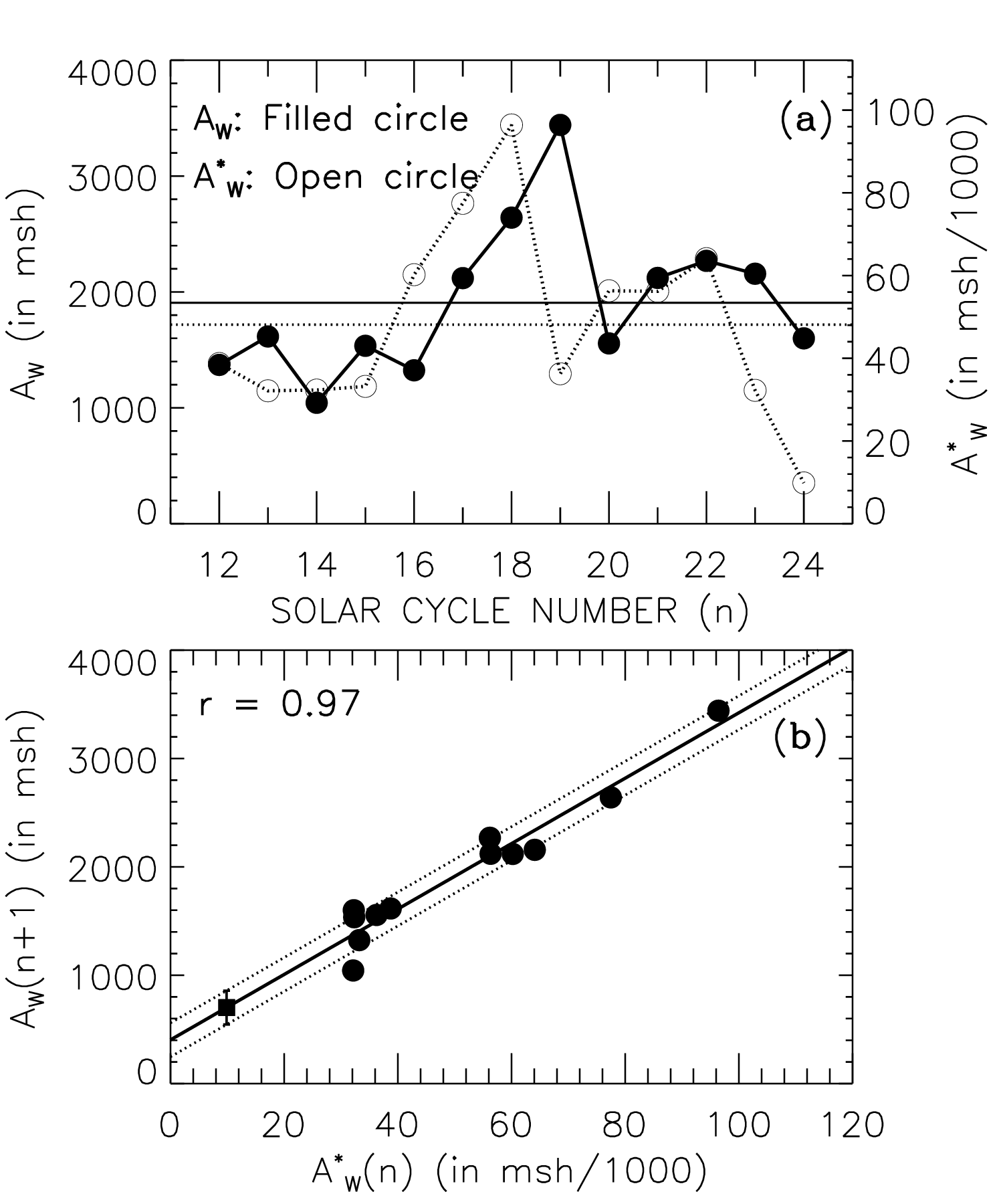}
\caption{(a) Plot of   $A_{\rm W}$, $i.e.$ the 13-month smoothed 
monthly mean  area of the  
sunspot groups in the Sun's whole-sphere at the  maximum epoch of the 
  solar cycle ({\large $\bullet{\rm-}\bullet$} {\it curve})-- and 
$A^*_{\rm W}$,  $i.e.$ the sum of the areas of the sunspot groups  
in $0^\circ - 10^\circ$ latitude interval of the southern hemisphere 
during the interval $T^*_{\rm W}$ just after  $T_{\rm M}$
of a solar cycle ({\bf \tiny $\bigcirc\cdots\bigcirc$} 
{\it curve})--versus the solar cycle number.
(b) The scatter plot of $A^*_{\rm W}$ of a Solar Cycle $n$  versus 
 $A_{\rm W}$  of  Solar Cycle $n+1$. The {\it continuous line}  
represents  the corresponding  linear best-fit and  
 the {\it dotted lines} are drawn at one-rmsd level. 
  The values of the correlation coefficient ($r$) is also given.
 The $\blacksquare$ represents 
the predicted value $701 \pm 156$ for $A_{\rm W}$ of Cycle~25.}
\label{fig3}
\end{figure}

\begin{figure}[ht]
\centering
\includegraphics[width=8.0cm]{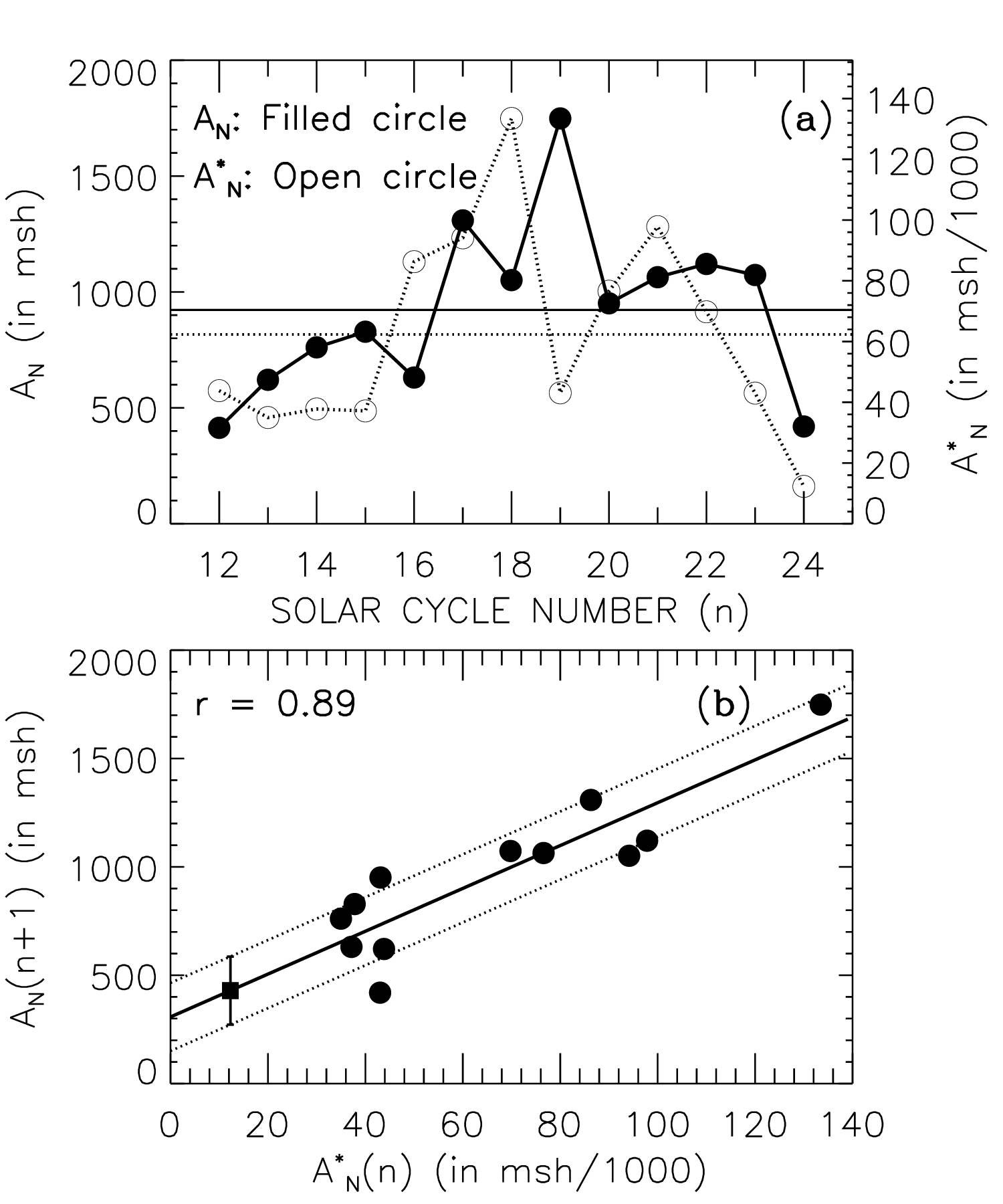}
\caption{(a) Plot of   $A_{\rm N}$, $i.e.$ the 13-month smoothed
 monthly mean  area of the 
sunspot groups in the northern hemisphere at the  maximum epoch of the 
  solar cycle ({\large $\bullet{\rm-}\bullet$}{ \it curve})--and
$A^*_{\rm N}$, 
 $i.e.$ the sum of the areas of the sunspot groups  
in $0^\circ - 10^\circ$ latitude interval of the southern hemisphere 
during the interval $T^*_{\rm N}$ just after  $T_{\rm M}$  
of a solar cycle ({\bf \tiny $\bigcirc\cdots\bigcirc$} {\it curve})--versus
 the solar cycle number.
(b) The scatter plot of $A^*_{\rm N}$ of a Solar Cycle $n$  versus 
 $A_{\rm N}$  of  Solar Cycle $n+1$. The {\it continuous line}  
represents  the corresponding linear best-fit and  
 the {\it dotted lines}  are drawn at one-rmsd level. 
   The values of the correlation coefficient ($r$) is also given. 
 The $\blacksquare$ represents 
the predicted value $429 \pm 157$ for $A_{\rm N}$ of Cycle~25.}
\label{fig4}
\end{figure}

\begin{figure}[ht]
\centering
\includegraphics[width=8.0cm]{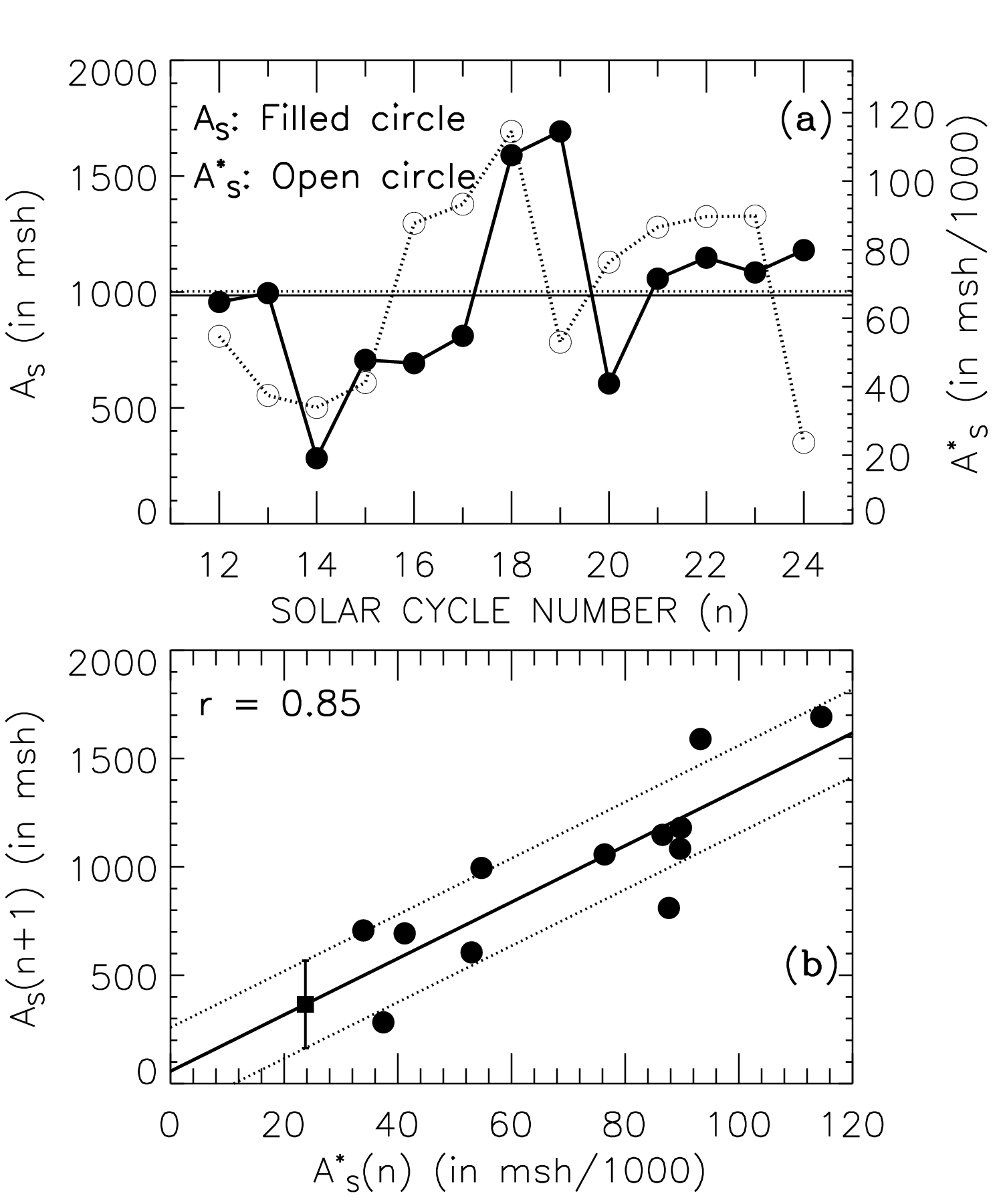}
\caption{(a) Plot of  $A_{\rm S}$, $i.e.$ the 13-month smoothed monthly
 mean area of the  
sunspot groups in the southern hemisphere at the  maximum epoch of the 
  solar cycle ({\large $\bullet{\rm-}\bullet$} {\it curve})--and $A^*_{\rm S}$, 
 $i.e.$ the sum of the areas of the sunspot groups  
in $0^\circ - 10^\circ$ latitude interval of the southern hemisphere 
during the interval $T^*_{\rm S}$ just after  $T_{\rm M}$  
of a solar cycle ({\bf \tiny $\bigcirc\cdots\bigcirc$} 
{\it curve})--versus the solar cycle number.
(b) The scatter plot of $A^*_{\rm S}$ of a Solar Cycle $n$  versus 
 $A_{\rm S}$  of  Solar Cycle $n+1$. The {\it continuous line}  
represents  the corresponding linear best-fit and  
 the {\it dotted lines}  are drawn at one-rmsd level.
  The values of the correlation coefficient ($r$) is also given. 
 The $\blacksquare$ represents 
 the predicted value $366 \pm 202$ for $A_{\rm S}$ of Cycle~25.}
\label{fig5}
\end{figure}

\begin{figure}[ht]
\centering
\includegraphics[width=8.0cm]{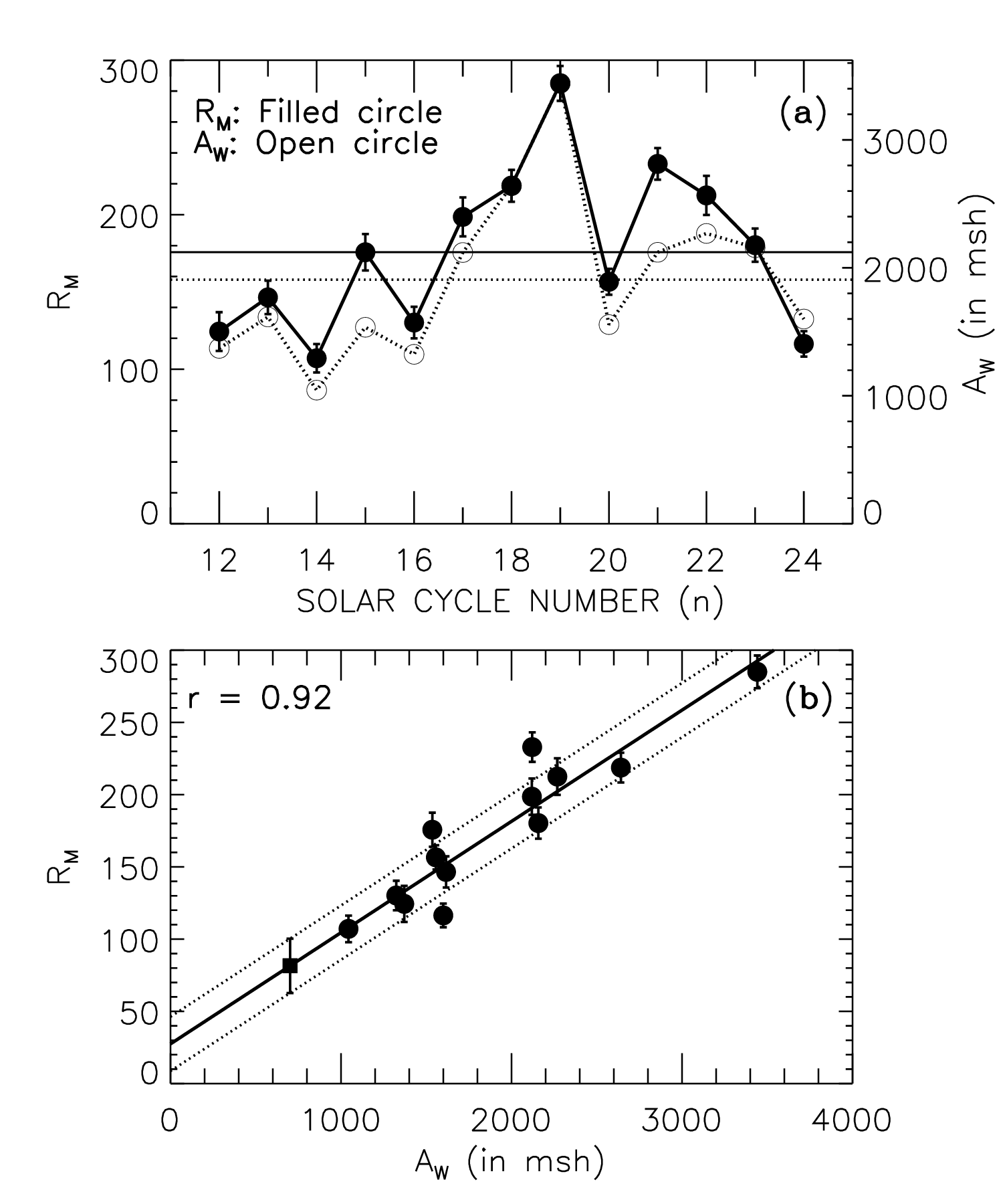}
\caption{(a) Plot of $A_{\rm W}$, 
   $i.e.$ the 13-month smoothed monthly mean  area of the
sunspot groups in the whole-sphere at the  maximum epoch of a 
  solar cycle ({\bf \tiny $\bigcirc\cdots\bigcirc$} {\it curve})--and 
 the amplitude  ($R_{\rm M}$) of
 the solar cycle ({\large $\bullet{\rm-}\bullet$} {\it curve})--versus the 
solar cycle number.
(b) The scatter plot of $A_{\rm W}$   versus 
 $R_{\rm M}$. The {\it continuous line}  
represents  the corresponding  linear best-fit and  
 the {\it dotted lines}  are drawn at one-rmsd level.
  The values of the correlation coefficient ($r$) is also shown.
 The $\blacksquare$ represents 
the corresponding predicted value $81 \pm 19$ for $R_{\rm M}$ of
 Cycle~25.}
\label{fig6}
\end{figure}

In Figs.~\ref{fig2}(a), \ref{fig3}(a), \ref{fig4}(a), and \ref{fig5}(a) we
 have compared the variations in 
$A^*_{\rm R}$, $A^*_{\rm W}$,   $A^*_{\rm N}$, and  $A^*_{\rm S}$  
 with the variations in   $R_{\rm M}$,  $A_{\rm W}$,  
  $A_{\rm N}$, and  $A_{\rm S}$, respectively, 
during  Solar Cycles 
12\,--\,24. As can be seen in these figures each of the former 
parameters leads the corresponding one of the latter parameters by about 
one-cycle period. Figs.~\ref{fig2}(b), \ref{fig3}(b), \ref{fig4}(b),
 and \ref{fig5}(b) show the     
 correlations of  
$A^*_{\rm R}$, $A^*_{\rm W}$,   $A^*_{\rm N}$, and  $A^*_{\rm S}$  of
  Cycle $n$  with    $R_{\rm M}$,  $A_{\rm W}$,  
  $A_{\rm N}$, and  $A_{\rm S}$, respectively, of Cycle $n+1$. 
 The values 0.94, 0.97, 0.89 and  0.85 of the corresponding 
correlation coefficients are also shown. 
 All of these  correlations are significant on $>$99.9\,\% confidence level 
($P < 0.001$).  
(The corresponding Student's $t$ values are  8.5, 12, 6,
    and 5.1 respectively. In the case of  
eleven degrees of freedom $t = 4.44$  for $P = 0.001$.)
We  obtained the following linear relationships.
\begin{equation}
\label{eqn1}
 R_{\rm M} (n+1)=  (1.98 \pm 0.12) A^*_{\rm R} (n) + 74 \pm 7 ,
\end{equation} 
\begin{equation}
\label{eqn2}
 A_{\rm W} (n+1) =  (30.2 \pm 2.5) A^*_{\rm W} (n) + 404 \pm 136 , 
\end{equation} 
\begin{equation}
\label{eqn3}
 A_{\rm N} (n+1)=  (9.9 \pm 1.6) A^*_{\rm N} (n) + 308 \pm 120 , 
\end{equation} 
and
\begin{equation}
\label{eqn4}
 A_{\rm S} (n+1) =  (13.0 \pm 2.5) A^*_{\rm S} (n) + 57 \pm 192 . 
\end{equation} 
 The fit of  each of these equations to 
the corresponding data  is reasonably good.  The  slopes of  
({\bf \ref{eqn1}}), ({\bf \ref{eqn2}}), ({\bf \ref{eqn3}}), and
 ({\bf \ref{eqn4}}) are about  
16, 12, 6, and 5 times larger than the corresponding uncertainties, 
respectively. 
In each of these cases, 
except one or two data points,  all of the remaining  data points are inside 
or close to the 
one-rmsd ({\it root-mean-square deviation}) level
 (see Figs.~\ref{fig2}\,--\,\ref{fig5}). Since 
errors ($\sigma_{\rm _R}$) in the values of $R_{\rm M}$ are available
 (see Table~\ref{table1}), 
hence, they are taken into account  in the fit of  {\bf \ref{eqn1}}. 
That is, the least-square fit is weighted with
 ${\rm weight} = \frac{1}{\sigma^2_{\rm _R}}$.
 Errors in the areas of  sunspot groups are not available.  
Using ({\bf \ref{eqn1}})\,--\,({\bf \ref{eqn4}}) and the values 
   of $A^*_{\rm R}$,  $A^*_{\rm W}$,   $A^*_{\rm N}$, and  $A^*_{\rm S}$ 
 of  Cycle~24 that are given in Table~\ref{table1},
 we obtained $86\pm18$, $701\pm156$ msh, $429\pm157$ msh, 
and $366 \pm 202$ msh for   $R_{\rm M}$, 
 $A_{\rm W}$,   $A_{\rm N}$, and   $A_{\rm S}$, respectively, of 
 Cycle 25. The rmsd values look to be substantially 
large for the predicted values, because the ranges of $A_{\rm W}$,  
 $A_{\rm N}$, and  $A_{\rm S}$ values are large and the corresponding
 predicted values are close  to the lower limits of the ranges.
In Figs. 2(b), 3(b), 4(b), and 5(b) we have also shown the  predicted values. 
 These predicted values of $R_{\rm M}$, $A_{\rm W}$, and 
$A_{\rm S}$ are about $\approx$25\,\%, $\approx$56\,\%, and $\approx$69\,\%,
 respectively,   smaller than the corresponding observed values of
 Cycle~24, whereas
 the predicted value of $A_{\rm N}$ of Cycle~25 is almost equal 
to the corresponding observed value  of Cycle~24.     
 These  predicted  values of     $A_{\rm N}$ and   $A_{\rm S}$ 
are almost equal. 
 Thus, the corresponding  
 north--south asymmetry  is predicted to be only $0.08$. That is, 
there will be no significant north--south asymmetry in  $A_{\rm W}$ 
of the upcoming Solar Cycle~25.  
The sum,  $\approx$795 msh, of the   predicted  values of  
   $A_{\rm N}$ and   $A_{\rm S}$
 is slightly larger than  the predicted value of $A_{\rm W}$, 
but the difference between this sum and the predicted value of $A_{\rm W}$ 
is within one-rmsd level of the latter.
The  predicted values, particularly, in the cases of   
$A_{\rm N}$ and   $A_{\rm S}$, have a very large uncertainties 
(the values of  rmsd are large). However, the corresponding linear 
best-fits are
 reasonably good  because
 the corresponding 
slopes are about 5\,--\,6 times larger than their respective
 standard deviations.  The corresponding north--south asymmetry is 
reasonably accurate because the uncertainty in 
the ratio of two parameters is much smaller than the 
corresponding uncertainties of the individual parameters 
\citep[see][]{jg97}.

 The values  of   $A^*_{\rm W}$,    $A^*_{\rm N}$, and  
  $A^*_{\rm S}$ of Cycle~24 are substantially   (a factor of three times)
smaller than the corresponding 
values  of all the  previous cycles (see Table~\ref{table1}). 
 This  is  consistent with that  Cycle~24 is a small cycle and moreover, 
 in the decline phase of this cycle the  southern hemisphere's activity  is  
 decreased more steeply than in the corresponding phases of   other cycles 
\citep[see Figure~1 in][also see Figure~\ref{fig10} below]{jj20}.  
 As already mentioned above, the epochs $T^*_{\rm W}$, 
 $T^*_{\rm N}$, and  $T^*_{\rm S}$  are close to the time of change
 in polarities  of global magnetic fields.  In  our earlier articles, 
\citep{jj08,jj15},  we have suggested equatorial crossing of magnetic 
flux/down flows could be a possible physical process behind  all the
 relationships, similar to ({\bf \ref{eqn1}}) above, found there.
During  Cycle~24 meridional motions 
of sunspot groups were found to be large and mostly  equatorward 
and poleward directions in  the northern- and southern-hemispheres, 
respectively \citep[see Figure~1 in][]{jj18}. In the text of the
earlier article \cite{jj18} 
 the directions of the meridional motions  were incorrectly mentioned 
as equatorward in  both the hemispheres, we regret for that mistake. 
The meridional motions 
of sunspot groups may depict some extent the meridional plasma flows in 
the solar convection zone. 
Therefore, we suggest that due to the aforementioned 
 south-bound  meridional flows there might be 
 equatorial crossing of a large  northern-hemisphere magnetic flux and   
  a large cancellation of southern-hemisphere magnetic flux.
This might contributed to Cycle~24 being  weaker than previous cycles  and 
 the values of $A^*_{\rm W}$, $A^*_{\rm N}$, and $A^*_{\rm S}$ being smaller 
in Cycle~24 than  in previous cycles.
 However, sustained cross-equatorial meridional flows are ruled out by dynamical
and symmetry considerations \citep{nort14}. Therefore, the  physical 
reason behind the small values of the aforementioned parameters in Cycle~24 
yet to be found.

\begin{figure}[ht]
\centering
\includegraphics[width=8.0cm]{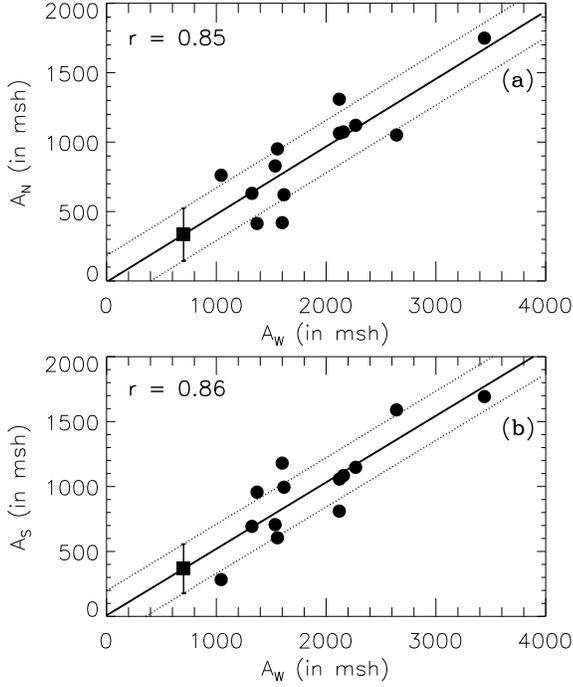}
\caption{The scatter plots of (a) $A_{\rm W}$   versus  $A_{\rm N}$ and (b) 
  $A_{\rm W}$   versus  $A_{\rm S}$.  
 The {\it continuous line}  
represents  the corresponding linear  best-fit and  
 the {\it dotted lines}  are drawn at one-rmsd level.
   The values of the correlation coefficient ($r$) is also shown.
 The $\blacksquare$ represents 
the corresponding predicted value $334 \pm 190$ for $A_{\rm N}$ and
 $467 \pm 190$ for $A_{\rm S}$  of Cycle~25.}
\label{fig7}
\end{figure}

\begin{figure}[ht]
\centering
\includegraphics[width=8.0cm]{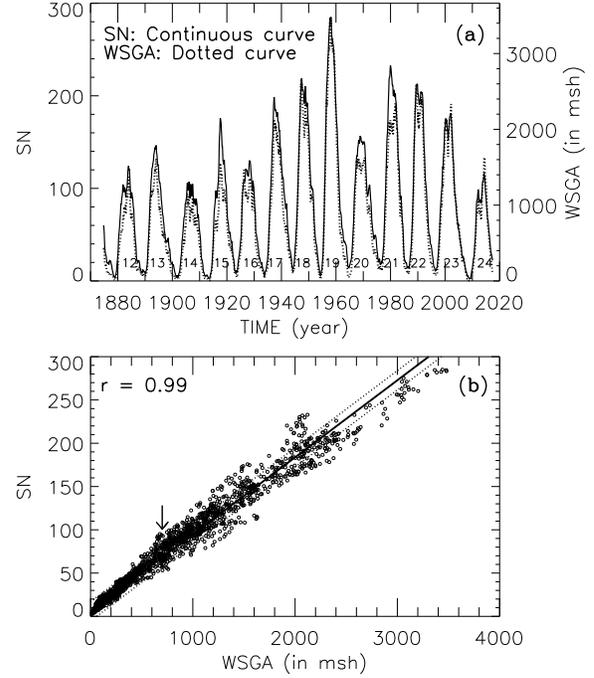}
\caption{(a) Plot of SN ({\it continuous curve}) and 
 WSGA ({\it dotted curve})    $i.e.$ 
the 13-month smoothed monthly mean values of total sunspot number (ISSN) and
   sunspot-group area in whole-sphere, respectively, versus the 
time (middle month of each 13-month interval) during the period 1875\,--\,2017.
(b) The scatter plot of WSGA   versus 
 $SN$. The {\it continuous line}  
represents  the corresponding  linear best-fit and  
 the {\it dotted lines}  are drawn at one-rmsd level.
  The values of the correlation coefficient ($r$) is also shown.
 The $\blacksquare$ (below $\downarrow$) at (701, 68) represents 
the corresponding predicted value $68 \pm 11$ for $R_{\rm M}$ of Cycle~25.}
\label{fig8}
\end{figure}

\begin{figure}[ht]
\centering
\includegraphics[width=8.0cm]{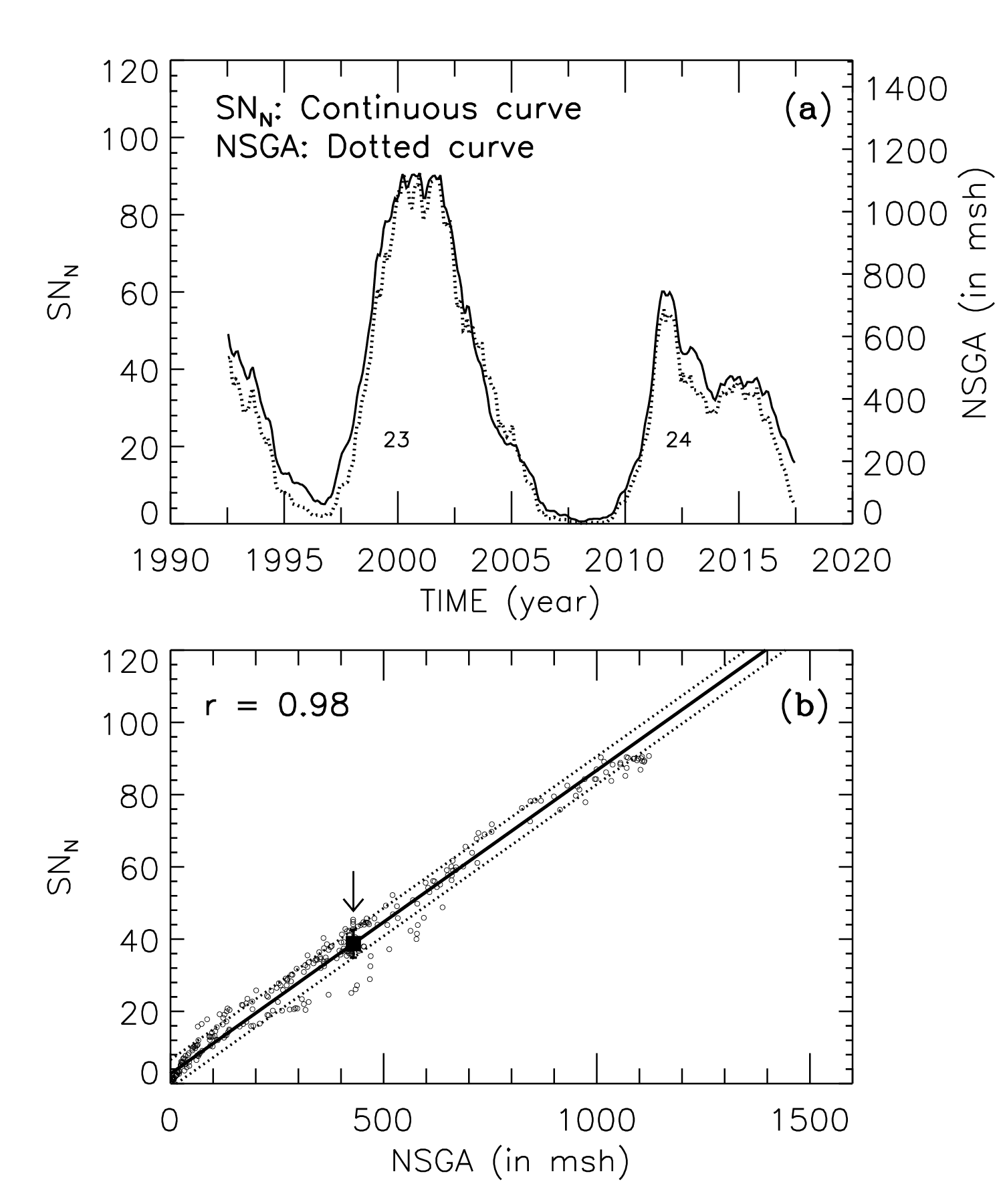}
\caption{(a) Plot of ${\rm SN}_{\rm N}$ ({\it continuous curve}) and 
  NSGA ({\it dotted curve}),  
   $i.e.$ the 13-month smoothed monthly mean values of ISSN  and 
sunspot-group area, respectively, in 
  northern-hemisphere  
  versus  time (middle month of each 13-month interval) during the
 period 1992\,--\,2017.
(b) The scatter plot of NSGA  versus 
 $SN_{\rm N}$. The {\it continuous line}  
represents  the corresponding  linear best-fit and  
 the {\it dotted lines}  are drawn at one-rmsd level.
  The values of the correlation coefficient ($r$) is also shown.
 The $\blacksquare$ (below $\downarrow$) at (429, 39) represents 
the corresponding predicted value $39 \pm 4$ for $SN_{\rm N}$  at 
the maximum epoch of Cycle~25.}
\label{fig9}
\end{figure}

\begin{figure}[ht]
\centering
\includegraphics[width=8.0cm]{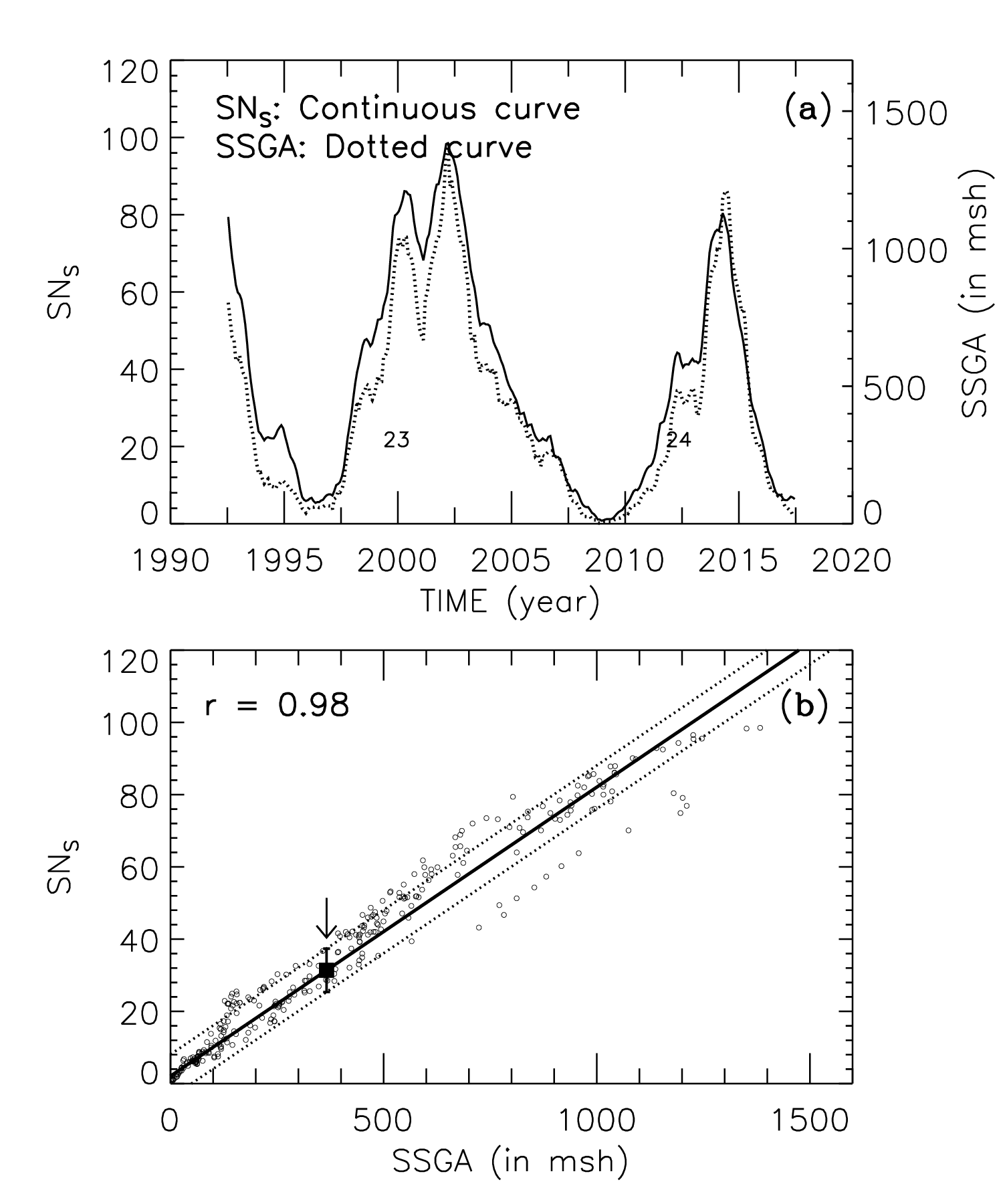}
\caption{(a) Plot of ${\rm SN}_{\rm S}$ ({\it continuous curve}) and 
SSGA ({\it dotted curve}), $i.e.$ the 13-month smoothed monthly
 mean values of  ISSN and 
 sunspot-group area, respectively, in southern-hemisphere versus
 time (middle month of a 13-month interval) during the period 1875\,--\,2017.
(b) The scatter plot of SSGA  versus 
 $SN_{\rm S}$. The {\it continuous line}  
represents  the corresponding  linear best-fit and  
 the {\it dotted lines}  are drawn at one-rmsd level.
  The values of the correlation coefficient ($r$) is also shown.
 The $\blacksquare$ (below $\downarrow$) at (366, 31) represents 
the corresponding predicted value $31 \pm 6$ for $SN_{\rm S}$ at the
 maximum epoch of Cycle~25.}
\label{fig10}
\end{figure}

\begin{table}[ht]
\centering
\caption{The values predicted for various parameters of Solar Cycle~25.}
\begin{tabular}{lllc}
\hline
  \noalign{\smallskip}
Equation&Relation between&Predicted\\
  \noalign{\smallskip}
\hline
  \noalign{\smallskip}
({\bf \ref{eqn1}})&$A^*_{\rm R}$ (n) and $R_{\rm M}$ (n+1)& $R_{\rm M} = 86 \pm 18$\\ 
({\bf \ref{eqn2}})&$A^*_{\rm W}$ (n) and $A_{\rm W}$ (n+1)& $A_{\rm W}  \approx 701$ \\
({\bf \ref{eqn3}})&$A^*_{\rm N}$ (n) and $A_{\rm N}$ (n+1)& $A_{\rm N}  \approx 429$ \\
({\bf \ref{eqn4}})&$A^*_{\rm S}$ (n) and $A_{\rm S}$ (n+1)& $A_{\rm S}  \approx 366$ \\
({\bf \ref{eqn5}})&$A_{\rm W}$ and $R_{\rm M}$& $R_{\rm M} = 84 \pm 19$ \\
      -        &$A_{\rm W}$ and $A_{\rm N}$& $A_{\rm N} \approx 334 $ \\
      -        &$A_{\rm W}$ and $A_{\rm S}$& $A_{\rm S}  \approx 467 $ \\
({\bf \ref{eqn6}})&WSGA and ${\rm SN}$ & $R_{\rm M} = 68 \pm 11$ \\
({\bf \ref{eqn7}})&NSGA and ${\rm SN}_{\rm N}$&${\rm SN}_{\rm N} = 39 \pm 4$\\ 
({\bf \ref{eqn8}})&SSGA and ${\rm SN}_{\rm S}$&${\rm SN}_{\rm S} = 31 \pm 6$ \\
\hline
  \noalign{\smallskip}
\end{tabular}
\label{table2}
\end{table}

\begin{figure}[ht]
\centering
\includegraphics[width=8.0cm]{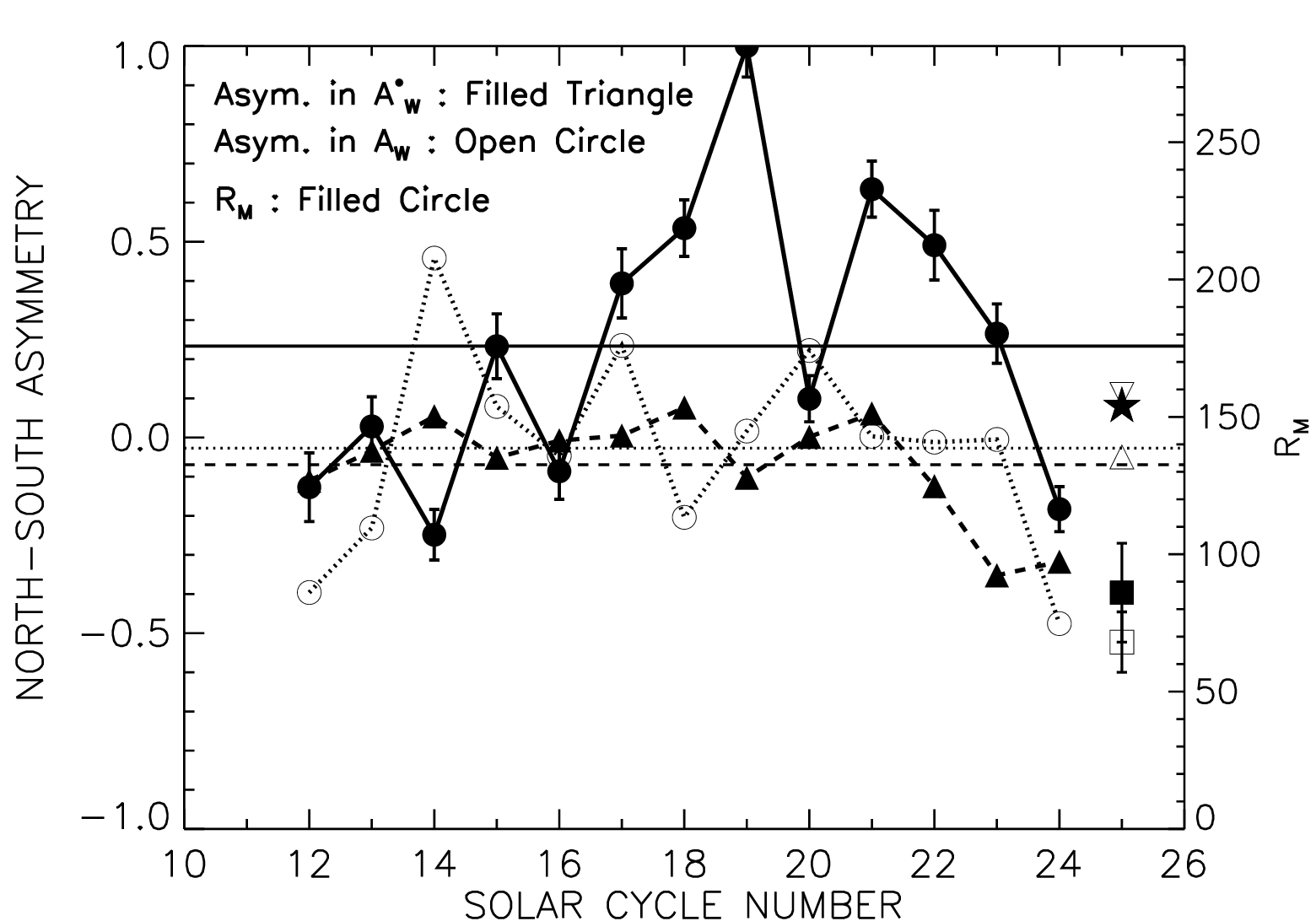}
\caption{Variations  in the corresponding 
 north--south asymmetry of $A_{\rm W}$ 
({\bf \tiny $\bigcirc\cdots\bigcirc$} {\it curve}),  
$i.e.$ in $\frac{A_{\rm N} - A_{\rm S}}{A_{\rm N} + A_{\rm  S}}$ and of
$A^*_{\rm W}$ ($\blacktriangle${\rm - -}$\blacktriangle$ {\it curve}),
 $i.e.$ in   
$\frac{A^*_{\rm N} - A^*_{\rm S}}{A^*_{\rm N} + A^*_{\rm  S}}$ during 
Cycles 12\,--\,24.  The {\large $\bullet{\rm-}\bullet$}  
{\it curve} represents the variation in $R_{\rm M}$. 
The $\blacksquare$  and {\bf $\square$} represent the  
values $86 \pm 18$ and $68 \pm 11$ of  $R_{\rm M}$ of Cycle~25 predicted from 
{\bf (\ref{eqn1})} and {\bf (\ref{eqn6})}, respectively, and 
the {\bf $\triangledown$} represents the value 0.11 of
  north--south asymmetry  in $R_{\rm M}$ predicted by using
{\bf (\ref{eqn7})} and {\bf (\ref{eqn8})}.
The  $\bigstar$ represents  the  value 0.08  of  the corresponding 
 north--south asymmetry in $A_{\rm W}$ of Cycle~25 predicted by using 
{\bf (\ref{eqn3})} and {\bf (\ref{eqn4})}, and the value $-0.05$ that 
  predicted by using the relations shown in  Fig.~\ref{fig7} is represented by 
 {\bf $\triangle$}. A positive (negative) value of the
 north--south asymmetry indicates the northern (southern) hemisphere's 
dominance.} 
\label{fig11}
\end{figure}

We also determined the correlation between $R_{\rm M}$ and $A_{\rm W}$. It is  
 shown in Fig.~\ref{fig6}. We found  $r = 0.92$ (the corresponding 
Student's $t = 7.8$) 
and obtained the following linear relationship:
\begin{equation}
\label{eqn5}
 R_{\rm M}  =  (0.077 \pm 0.005) A_{\rm W} + 27.5 \pm 9.3 .  
\end{equation}
The fit of this equation to the data is very good (in this fit error in  
$R_{\rm M}$ is taken into account). The slope is 15 times larger than  
the corresponding standard deviation.  By using the value
 701 msh  predicted  for $A_{\rm W}$ of 
Cycle~25 above, we have  obtained $81\pm 19$ 
for  $R_{\rm M}$ of this cycle, 
 which is almost  same as the value predicted  by using ({\bf \ref{eqn1}})
 above. Thus, overall, there is a suggestion that the amplitude  of
 Cycle~25 will
 be some extent smaller than that of Cycle~24.

As already mentioned  above, both  
 $A_{\rm N}$ and $A_{\rm S}$ reasonably well correlate with 
 $A_{\rm W}$.  By using these relationships that are shown 
in Fig.~\ref{fig7} and the value 701 msh of $A_{\rm W}$ of Cycle~25 that
 is predicted above, we obtained the values $\approx$334 msh and
 $\approx$467 msh  for $A_{\rm N}$ and $A_{\rm S}$, respectively, of Cycle~25.
From these values  also we get north--south asymmetry in $A_{\rm W}$ of
 Cycle 25 is small ($-0.05$), but it is a small  negative value 
(indicates a slight more activity in south).

In our earlier analyzes, \citep{jj07,jj08}, we have also found that
 the sum (here namely $a^*_{\rm R}$) of the areas of the  
sunspot groups in $0^\circ - 10^\circ$ latitude interval of the northern 
hemisphere--and in the time interval (namely $T^*_{\rm m}$) $- 1.35$ year to
 $+ 2.15$ year from the time of preceding minimum ($T_{\rm m}$)--of Cycle $n$ 
 well-correlate with the $R_{\rm M}$ of Cycle $n+1$. 
Although by using this relationship also we have
 found $R_{\rm M}$ of Cycle 24 is 
slightly smaller than that of Cycle 23, but it is found to be considerably
 higher 
than the observed $R_{\rm M}$ of Cycle~24. That is,  this relationship in
 general  yields  a considerable over estimate for the amplitude of a cycle.  
From this method here we  got slightly different values for 
$T^*_{\rm m}$. We obtained $107 \pm 25$ for $R_{\rm M}$ of 
Cycle 25, and the values $1223 \pm  233$ msh, $619 \pm 144$ msh, and
 $608 \pm 231$ msh for $A_{\rm W}$, $A_{\rm N}$, and $A_{\rm S}$ of
 Cycle 25, respectively.  The corresponding correlations are 
considerably smaller than those of  
({\bf \ref{eqn1}})\,--\,({\bf \ref{eqn4}}), above.  
 These values of $A_{\rm N}$ and $A_{\rm S}$ are also
  almost equal. That is, from this method also we obtain a negligibly
small  value (0.009) for the north--south asymmetry in $A_{\rm W}$  of
 Cycle~25. We would like to mention that the value of $R_{\rm M}$ of Cycle~24 
predicted by us~\citep{jj07} using the method that correspond to 
({\bf \ref{eqn1}}) is found about 10\,\% smaller than the observed
 $R_{\rm M}$. However, in that earlier analysis 
we have used Solar Optical Observing Network (SOON) sunspot group data 
for Cycle~21\,--\,23, in which the areas of  sunspot groups seem to be 
some extent under estimated even after increased by a factor of 1.4
\citep{hath08}.  There is no such problem in the DPD data. Hence,   
 the values predicted here  for the parameters of Cycle~25 by using 
 ({\bf \ref{eqn1}})- ({\bf \ref{eqn4}}) above,  may be  not  under 
estimated , $i.e.$  may not be about 10\,\% smaller
 than the respective observed ones. 
  Nevertheless, our predictions indicate that
 Cycle~25 will be at least few percent weaker than Cycle~24.

Here we would like to point out that  as already mentioned above  
 $R_{\rm M}$ of Cycle 21 is slightly larger than that of Cycle 22, whereas 
the behavior of $A_{\rm W}$ seems opposite to this (see Fig.~\ref{fig1}). 
However, still  a reasonably high correlation exits between  $R_{\rm M}$ 
and $A_{\rm W}$ ($cf.$, Fig.~\ref{fig6}). We have also determined the 
correlation between  the 13-month smoothed monthly mean   
 sunspot-group area and ISSN in whole sphere (/total), northern hemisphere,
 and   southern hemisphere,  $i.e.$ the correlations between  WSGA and SN,   
NSGA and ${\rm SN_{\rm N}}$, and SSGA and ${\rm SN_{\rm S}}$, respectively.
The whole sphere  has 1713 SN-values  
during October 1874\,--\,June 2017, whereas the northern and southern
 hemispheres have 300 ${\rm SN_{\rm N}}$- and ${\rm SN_{\rm S}}$-values
 during July 1992\,--\,June 2017.
Figs.~8, 9, and 10 show the corresponding
 relations in the whole sphere,  northern hemisphere, and 
 southern hemisphere, respectively. In these figures the corresponding values
  0.99, 0.98, and 0.98 of $r$ are also shown. Each of these values of $r$
 is vary high \citep[also see][]{hath02} and  
 obviously, the corresponding linear best-fit is  
 very good. We obtained the following relations:
\begin{equation}
\label{eqn6}
{\rm SN} =  (0.089 \pm 0.0002) {\rm WSGA} + 5.74 \pm 0.15 ,
\end{equation}
\begin{equation}
\label{eqn7}
{\rm SN}_{\rm N} =  (0.084 \pm 0.0008) {\rm NSGA} + 2.72 \pm 0.19 , 
\end{equation}
and
\begin{equation}
\label{eqn8}
{\rm SN}_{\rm S} =  (0.08 \pm 0.0007 ) {\rm SSGA} + 2.08 \pm 0.18 .
\end{equation}
In the fit of each of these equations the errors in the values of ISSN
are taken into account. 
Using ({\bf \ref{eqn6}}), ({\bf \ref{eqn7}}), and
 ({\bf \ref{eqn8}}) and the  values of 
$A_{\rm W}$, $A_{\rm N}$, and $A_{\rm S}$ that are 
predicted  from  ({\bf \ref{eqn2}})\,--\,({\bf \ref{eqn4}}) above 
($i.e$ the values of 
WSGA, NSGA, and SSAG  at the epoch of   
 $R_{\rm M}$ of Cycle~25),    
we obtained $68 \pm 11$  for
 $R_{\rm M}$ of Cycle~25   and  
$39 \pm 4$ and $31 \pm 6$ for the values 
of ${\rm SN_{\rm N}}$ and ${\rm SN_{\rm S}}$
 at the epoch of  $R_{\rm M}$ of Cycle~25.  It may be noted here 
 these predicted values 
of ${\rm SN_{\rm N}}$ and ${\rm SN_{\rm S}}$ may or may not represent 
 the values of  northern- and southern-hemispheres' sunspot 
peaks of Cycle~25.
 This predicted value of $R_{\rm M}$ of Cycle~25 is
 some extent smaller than that predicted by using ({\bf \ref{eqn1}})
 above, and about 40\,\% smaller 
than that of Cycle~24.  This predicted
 value of $R_{\rm M}$ of Cycle 25  may be
 more accurate than that predicted from ({\bf \ref{eqn1}}) because   
the corresponding correlation of ({\bf \ref{eqn2}}) 
is higher than that of ({\bf \ref{eqn1}}). That is, 
the value of  $A_{\rm W}$ of Cycle~25 used in ({\bf \ref{eqn6}})
 is more accurately predicted from
({\bf \ref{eqn2}})  than that of $R_{\rm M}$ of Cycle~25 predicted from 
({\bf \ref{eqn1}}). 
However, as can be seen in Figs. 8(a), 9(a), and 10(a)  
the variations in  
 ISSN and sunspot-group area  seem to be matching 
much better around the minimum period than around 
the maximum period of a cycle and
it should also be noted that Waldmeier effect is not present 
in the cycles of sunspot-group area \citep{dgt08,jj19}. 
The predicted value of ${\rm SN_{\rm N}}$ is about 20\,\% larger 
than that of  ${\rm SN_{\rm S}}$.  However, still there is a 
suggestion that the corresponding north--south  asymmetry (0.11) of
 $R_{\rm M}$ of Solar Cycle~25 
 will be much smaller than the absolute value of the north--south asymmetry 
($-0.38$) of $R_{\rm M}$ of Solar Cycle~24. Note that  
 at the epoch of   $R_{\rm M}$ of Cycle~24 
 the observed values of ${\rm SN}_{\rm N}$  and  ${\rm SN}_{\rm S}$ 
are  36.0, and 80.4, respectively.

In Table~\ref{table2} we have 
given the values predicted for various parameters of Solar Cycle~25.
 Fig.~\ref{fig11} shows the cycle-to-cycle variation in the corresponding 
 north--south asymmetry of $A_{\rm W}$, $i.e.$ in  
$\frac{A_{\rm N} - A_{\rm S}}{A_{\rm N} + A_{\rm  S}}$ and 
 of   $A^*_{\rm W}$, $i.e.$ in 
$\frac{A^*_{\rm N} - A^*_{\rm S}}{A^*_{\rm N} + A^*_{\rm  S}}$ during  
Cycles 12\,--\,24. In this  figure we have also shown the cycle-to-cycle 
variation in $R_{\rm M}$. In addition,   the  predicted values of  
$R_{\rm M}$  and the corresponding north--south asymmetry of
$A_{\rm W}$ of Cycle~25 were also shown. 
There exist no significant 
correlations between  $R_{\rm M}$ and the corresponding north--south 
asymmetry of  either $A_{\rm W}$ or  $A^*_{\rm W}$. 
In addition, as can be seen in Fig.~\ref{fig11}, the corresponding 
 asymmetry of
 $A_{\rm W}$ of each of  Cycles~13,  17, and 20 has 
  a considerable positive value and that of each of Cycles~12, 13, and 24
  has a considerable negative value. There is a suggestion that  
the amplitude of Cycle~25 will  have  a positive or a negative very 
small (/negligible) north--south  asymmetry
  similar as that correspond to the average   
 Cycle~15 or  the small Cycle~16 (see Fig.~\ref{fig11}).
Overall, there is a suggestion that there will be no significant 
 north--south asymmetry correspond to the amplitude of Cycle~25.   

Note: we have repeated all the above calculations  by using Version-1
 of ISSN-series and obtained the 
 relations similar as ({\bf \ref{eqn1}})--({\bf \ref{eqn4}}), but  
 their  corresponding correlations   
 are found better than those of  ({\bf \ref{eqn1}})--({\bf \ref{eqn4}}).
 We found $51\pm 12$ for 
$R_{\rm M}$ (Version-1) of Cycle 25, which is about 38\,\% less than that 
(81.9) of Cycle 24 and the corresponding north--south asymmetry is also 
found negligible.

 As mentioned in  Section~1,  
 in some cycles  the northern- and southern-hemispheres typically peaks 
at different times, $i.e.$ they are out of phase. 
 At least one of the hemispheric maxima frequently doesn't coincide with 
the overall maximum. In some cycles 
northern-hemisphere's peak is strong and in some other cycles 
southern-hemisphere's peak is strong. The strong peak is always do not 
occur first in the same hemisphere~\citep{nort14,jj19}. 
In Figure~1 of our recent 
paper, \cite{jj20}, the properties such as dominant hemisphere, phase 
difference 
between the northern- and southern-hemispheres, {\it etc.} of Solar Cycles
 12--24   can be seen clearly. Regarding the relationships  
({\bf\ref{eqn1}})--({\bf \ref{eqn4}}), we 
would like to point out that in the case of  Cycle~24 its second SN-peak
 (coincided with the strong southern-hemisphere's peak)  is 
taken into account because  it is substantially larger than 
its first peak (coincided with the weak northern-hemisphere's peak),
 whereas in the case of most of the previous cycles the respective first peak
 is taken into account because it is larger than the corresponding 
second peak. Here we predicted the levels of activity in the  northern- and 
southern-hemispheres  at the epoch of the peak ($R_{\rm M}$)
 of whole-sphere activity of Cycle~25.
 However, the hemispheres may be somewhat independent 
\citep[$e.g.$][]{dg01,bd13}.  
Therefore, it may be necessary to make   a prediction for each hemisphere's
 maximum independently.

\begin{table*}[ht]
\centering
\caption{Predictions of Solar Cycle~25.}
\begin{tabular}{lllc}
\hline
  \noalign{\smallskip}
Authors&Prediction&Approach\\
  \noalign{\smallskip}
\hline
  \noalign{\smallskip}
{\bf Larger than Cycle~24:}\\
  \noalign{\smallskip}
\cite{hg12}&1.5 times of Cycle~24& From the statistics of spotless events.\\
\cite{jc17}& Stronger than Cycle~24& Surface flux transport model-based
method.\\
\cite{jiang18}& 93\,--\,155& Solar surface Flux transport dynamo model.\\ 
  \cite{sarp18}& $154 \pm 12$&  The empirical dynamical modeling method\\
&& to the revised Version-2 of ISSN series\\
&& starting from 1749 July to 2018 January.\\
\cite{ps18}& $135 \pm 25$& SODA Index determined by polar magnetic\\
&&fields and  spectral index as precursors.\\
\cite{gopal18}& 89 (south)& Used a precursor method \\
&59 (north)& (used microwave imaging observations).\\

\\
{\bf Similar to  Cycle~24:}\\
  \noalign{\smallskip}
\cite{du06}& $102.6 \pm 22.4$ & Maximum-maximum cycle length as an\\
&& indicator to predict the amplitude.\\
\cite{cameron16}& Not much higher& Expecting a value of dipole moment
 around 2020.\\
&than Cycle~24\\
\cite{wang17}& Similar to Cycle~24&Based on the observed evolution of polar
 fields\\
&& and the axial dipole component of the\\
&& Sun's global magnetic field at the end of 2015.\\ 
\cite{sb17}&$102.8\pm24.6$& A statistical test for persistence of solar\\
&& activity based on the value of Hurst exponent (H).\\
\cite{okoh18}& $122.1\pm 18.2$& Using the Ap-index as a precursor.\\
\cite{bn18}& Slightly stronger & Using magnetic field evolution\\ 
&than  Cycle~25& models for the Sun's surface and interior.\\
\cite{hawk18}& Similar as Cycle 24& By employing magnetic helicity\\
 &&as a predictor.\\
\cite{pishk19}& $116 \pm 12$&  Using  the maximal value of filtered Wilcox\\
 &&Solar Observatory polar field strength\\
&& before the cycle minimum as precursor.\\
\cite{miao20}& 121& Based on the linear regression relationship\\
&& between sunspot maxima and aa-index minima.\\

\\
{\bf Smaller than Cycle~24:}\\
  \noalign{\smallskip}
\cite{jbu05},& Current Gleissberg& A cosine-fit of amplitudes of\\
 \cite{jj17}&cycle's minimum& Cycles 1--24, {\it etc.}\\
  \noalign{\smallskip}
\cite{jj15}&  $72\pm14$ & The same as the method used in this article.\\
&(converted to Version~2)\\
  \noalign{\smallskip}
\cite{hath16}, &  A  weak Cycle~25& Using an  Advective Flux Transport code.\\
\cite{uh18}&(95\,\% of Cycle~24)\\ 
&South  more  than North\\
  \noalign{\smallskip}
\cite{lcl19} & $89^{+29}_{-14}$ & Data-driven solar cycle model.\\
&North 20\,\% more than South\\
&6 month onset delay in North\\
  \noalign{\smallskip}
\cite{cpf19}& $57 \pm 17$&  Based on the  feed-forward artificial\\
&& neural networks.\\
\cite{kit20}& $50\pm15$ (whole) & By using the Ensemble Kalman Filter\\
& $\approx 25$ (south)&method assimilated the poloidal and\\
& $\approx 23$ (north)& toroidal magnetic field components.\\
\hline
  \noalign{\smallskip}
\end{tabular}
\label{table3}
\end{table*}

\section{Conclusions and Discussion}
We have analyzed the daily sunspot-group  data  reported by 
 GPR during the period 1874\,--\,1976,  DPD  
 during the period 1977\,--\,2017, and the revised Version-2 of ISSN during 
the period 1874\,--\,2017. We determined the amplitudes and corresponding 
epochs of  Solar Cycles 12\,--\,24, and 
the 13-month smoothed monthly   
 mean corrected areas of the sunspot groups in the Sun's whole-sphere  
($A_{\rm W}$),   northern hemisphere ($A_{\rm N}$), and southern hemisphere
 ($A_{\rm S}$) at the epochs of the maxima of Solar Cycles 12\,--\,24.
By using all these  
 we obtained the relations--that are
 similar to the one found and used for predicting the amplitudes of 
Solar Cycles~24 and 25   in our  earlier 
analyzes~\citep{jj07,jj08,jj15}--separately for  the Sun's whole sphere
 and northern- and 
southern-hemispheres.  Using these relations  we predict the
values $\approx$701 msh, 
 $\approx$429 msh, and $\approx$366 msh   for  $A_{\rm W}$, 
 $A_{\rm N}$, and  $A_{\rm S}$, respectively, at the maximum epochs of 
the upcoming Solar Cycle 25. We  predict $86 \pm 18$ for  $R_{\rm M}$ 
of Solar Cycle~25. 
The 13-month smoothed monthly mean WSGA,
NSGA, and SSGA highly correlate with  SN, ${\rm SN}_{\rm N}$, and 
${\rm SN}_{\rm S}$, $i.e.$ the corresponding 
  total, northern-, and southern-hemispheres' sunspot
numbers, respectively. Using these relations and 
 the predicted values of $A_{\rm W}$,
 $A_{\rm N}$, and  $A_{\rm S}$ we obtain   
 $68 \pm 11$  for $R_{\rm M}$  of Solar Cycle~25, which 
is slightly lower than the aforementioned predicted value, and
$39 \pm 4$ and $31 \pm 6$ for the values of ${\rm SN}_{\rm N}$ and
  ${\rm SN}_{\rm S}$ at the  maximum of epoch of Solar Cycle~25. 
The aforementioned  predicted values of $R_{\rm M}$ of Solar Cycle~25 
 are 20\,\%--40\,\% smaller 
than the corresponding observed values of Solar Cycle~24.
The predicted values of $A_{\rm W}$, $A_{\rm S}$, and ${\rm SN}_{\rm S}$ 
  of Solar Cycle~25 are significantly smaller than the corresponding 
observed values of Solar Cycle~24, whereas the
predicted  values of $A_{\rm N}$ and ${\rm SN}_{\rm N}$ of Solar Cycle 25
 are almost equal to the
corresponding observed values of Solar Cycle 24.
The difference between the predicted  ${\rm SN}_{\rm N}$ and
 ${\rm SN}_{\rm S}$ and 
also between the predicted  $A_{\rm N}$ and  $A_{\rm S}$  
at the maximum epoch of Solar Cycle~25 are considerably small,
 $i.e.$ the corresponding values of the 
  north--south asymmetry  are very small (0.08\,--\,0.11).
In addition, $A_{\rm W}$ is 
found to be well correlated to both  $A_{\rm N}$ and   $A_{\rm S}$. From this
relationship also we get a  small value ($-0.05$) for the 
corresponding  north--south asymmetry 
in $A_{\rm W}$ of Solar Cycle~25. That is,  there is a  suggestion that 
there will be no significant north--south asymmetry in the amplitude
of the upcoming Solar Cycle~25. Overall, our results
suggest that   the amplitude of Solar
 Cycle~25 would be 25\,\%--40\,\% smaller,  
and the corresponding north--south asymmetry would be much smaller, 
 than those of Solar Cycle~24.

A number of  
 authors have provided several explanations  for the north--south asymmetry 
in the solar activity on both the  observational and
theoretical grounds \citep[for a detail review see][]{nort14}. 
Recent numerical simulations from a flux transport dynamo model
 show the meridional circulation works differently in the
 northern- and southern-hemispheres in producing differing solar cycles 
in the northern- and southern-hemispheres \citep{bd13}. 
The  flux-transport-dynamo process  
 is the physical mechanism behind our
 earlier \citep{jj07,jj08,jj15,jj19} and the present predictions. 
 All our predictions are consistent with the kind of 
flux transport models in which a long magnetic memory is an important 
criteria \citep[$e.g.$][]{dg06}.  As suggested in our earlier articles, 
here also we  suggest that the magnetic flux-transport by the solar meridional
 flows and  down-flows at active regions is the  main physical process 
behind our all predictions. 

 According to a kind of flux transport dynamo models 
the strength of the polar field at 
 the minimum of a cycle  would decide the strength of the same cycle.  
There exists a good correlation between the polar field
 at minimum epoch of cycle  and the amplitude  of the cycle~\citep{jcc07}. 
Hence,  the high correlation between
 the sum of the sunspot-group area in the southern-hemisphere 
near-equatorial band  during the small  
interval just after maximum of  Cycle~$n$  and the maximum of Cycle~$n + 1$, 
implies a high correlation between the former and the strength of the 
polar field at the following minimum ($i.e.$ at the preceding 
minimum of Cycle~$n + 1$). As pointed by one of the referees,  
 the values of $A^*_{\rm W}$,  $A^*_{\rm N}$, and $A^*_{\rm S}$ are all
 related to the same area on the Sun,
 but are so different (see Table~\ref{table1}) even though differences in the
 time intervals
 are very small. If the correlation actually indicates some kind of physical
 link, this would suggest that a very few active regions in this one narrow 
latitude band in the southern hemisphere over a relatively short time period 
are very important indicators of strength of the following cycle over many
 cycles. An interesting follow-up study would be to look in more detail at the
 cause of the differences. As already mentioned in the previous section, it 
might be worth thinking about cross-equatorial
 flux flows as well, which would be related to polar flux strength at the  
following minimum \citep{cameron14}. 

Earlier, a cosine-fit of the amplitudes of Sunspot Cycles~1--24 indicated  
 that Cycle~25 will take place at the minimum of the current Gleissberg cycle
\citep{jbu05,jj17}. This supports a low value
 predicted here for the amplitude of Sunspot Cycle~25.  
 Recently, \citep{jj19}, we predicted around May/2025 for  the maximum 
epoch of Cycle~25 of sunspot-group area (Area Cycle~25). 
Since no significant correlation between the rise-times 
of  sunspot- and  area-cycles was found, hence, in that article we did not 
 predict the maximum epoch of Sunspot Cycle~25.
 On the other hand,  
in Solar Cycles 22, 23, and 24--which are in  the 
descending phase of the current Gleissberg cycle--the highest peaks
 of ISSN coincided with those of sunspot-group area 
(see Figure~8). Hence,  the highest peaks of ISSN and area of  Cycle~25 
 may coincide. Therefore,  here  we predict 
around May/2025 also for  the  maximum epoch of ISSN Cycle~25.

Many authors predicted the amplitude of  Solar Cycle~25 by using
various methods: non-linear approaches, precursors,  dynamo 
 models, {\it etc.}  \citep[also see][]{sarp18,lcl19,pishk19,miao20}.  
In Table~\ref{table3} we have given a list of some of the earlier predictions. 
The recent international panel of experts coordinated by the NOAA and NASA,
 to which the WDC-SILSO contributed, reached a consensus
indicating that Cycle~25 will be most likely peak between 2023 and 2026 
with a maximum sunspot number between 95 and 130. The value 
we have predicted here for  the   
peak  of Sunspot Cycle~25 is close to (/slightly less than) the low value  
indicated by the penal.

\acknowledgments
The author thanks the anonymous referees of both the first and the revised 
versions of the manuscript for useful comments and  suggestions.
The author acknowledges the work of all the
 people contribute  and maintain the GPR and DPD  Sunspot databases.
The sunspot-number data are provided by WDC-SILSO, Royal Observatory of
Belgium, Brussels.

{}
\end{document}